\documentclass[a4paper,11pt]{article}

\usepackage[dvips]{graphicx}
\usepackage{amsmath,amssymb}
\usepackage{amsfonts}
\usepackage{epsfig}
\usepackage{subfigure,rotating,rotate}

\usepackage[affil-it]{authblk}

\newcommand{\Urfive}{$^{235}$U}
\newcommand{\Ureight}{$^{238}$U}
\newcommand{\Punine}{$^{239}$Pu}
\newcommand{\Puone}{$^{241}$Pu}
\newcommand{\neb}{$\bar{\nu}_{e} \; $}

\begin{document}

\title{The antineutrino energy structure in reactor experiments}

\author{Pau Novella\\
pau.novella@ific.uv.es}

\affil{Instituto de F\'isica Corpuscular (IFIC), CSIC and Universitat de Val\`encia\\
c/ Catedr\'atico Jos\'e Beltran 2, 46980 Paterna, Spain}

\date{ } 

\maketitle 

\begin{abstract}
The recent observation of an energy structure in the reactor antineutrino spectrum is reviewed. The reactor experiments Daya Bay, Double Chooz and RENO have reported a consistent excess of antineutrinos deviating from the flux predictions, with a local significance of about 4$\sigma$ between 4 and 6 MeV of the positron energy spectrum. The possible causes of the structure are analyzed in this work, along with the different experimental approaches developed to identify its origin. Considering the available data and results from the three experiments, the most likely explanation concerns the reactor flux predictions and the associated uncertainties. Therefore, the different current models are described and compared. The possible sources of incompleteness or inaccuracy of such models are discussed, as well as the experimental data required to improve their precision.
\end{abstract}

\section{Introduction}


In the last two decades, several neutrino oscillation experiments \cite{pdg} have demonstrated that neutrinos are massive particles. Thus, neutrinos have become a main probe to explore physics beyond the Standard Model. Within the three neutrino paradigm, the neutrino oscillation probability can be described by three mixing angles ($\theta_{12}$, $\theta_{23}$, $\theta_{13}$), two independent mass square differences ($\Delta m^2_{21}$, $\Delta m^2_{31}$), and one phase $\delta_{CP}$ responsible for the $CP$-violation in the leptonic sector. 
While the dominant oscillations driven by $\theta_{12}$ and $\theta_{23}$ have been measured by different experiments in the so-called solar and atmospheric sectors, the third mixing angle $\theta_{13}$ remained unrevealed until very recently. The first direct indications of a non-zero value of this angle has come from the accelerator-based experiments MINOS \cite{minos} and T2K \cite{t2k}. However, the current accelerator neutrino experiments cannot measure $\theta_{13}$ independently of other oscillation parameters. Complementing the role of accelerator-based facilities, reactor neutrino experiments stand as the direct way to provide an accurate value of $\theta_{13}$. In a two flavors scheme  and for short baselines (L$\sim$2km), the survival probability of a reactor electron anti-neutrino $\bar{\nu}_e$ with energy $E_\nu$ can be described as:

\begin{equation}
\label{eq:p}
P(\bar{\nu}_e \rightarrow \bar{\nu}_e) \cong 1
- \sin^2 2 \theta_{13} \sin^2\left(\frac{1.27\Delta m^2_{31}(\textrm{eV}^2) L(\textrm{m})}{E_{\nu}(\textrm{MeV})}\right),
\end{equation} 
  
\noindent The value of $\theta_{13}$ can be measured directly from the oscillation amplitude, inferred from an energy-dependent deficit in the number of observed neutrinos. 


After a series of short and medium-baseline ($\sim$100$-$1000 m) reactor neutrino experiments carried out in the 80s and 90s, a new generation of experiments has been operating for the past few years. Three different collaborations (Daya Bay \cite{db}, Double Chooz \cite{dcIII} and RENO \cite{reno}) have reported very precise measurements of the mixing angle $\theta_{13}$. As these experiments rely partially on the comparison of the observed antineutrino flux with respect to the expected one, a revision of the relatively old reactor flux models was performed in \cite{Mueller} and \cite{Huber}, becoming the new references and reducing the uncertainties at the 3\% level. This re-evaluation of the flux led to the so-called reactor antineutrino anomaly \cite{anom}, pointing at a possible short-baseline oscillation that would imply the existence of at least one sterile neutrino. While this suggests a possible underestimation of the reactor flux uncertainties, it does not impact the determination of $\theta_{13}$. Beyond this anomaly, Daya Bay, Double Chooz and RENO have reported very recently \cite{DBFlux,dcIII,RENONu14} an energy distortion around 5 MeV, which deviates from the expectation at about 3-4$\sigma$. Apart from reinforcing the idea of an underestimation of the flux errors, this experimental result has induced a world-wide effort in trying to understand the origin of this discrepancy.       


This work reviews the observation of such a 5 MeV energy structure by the three current reactor experiments. The possible causes are described as well as the different experimental approaches carried out to identify its origin. The incompleteness of the reactor flux predictions is presented as the most likely explanation, so the different models developed so far are reviewed. Within those models, a number of possible sources of biases or error underestimations are listed, thus pointing at possible experimental ways to improve our current knowledge. This review is organized as follows: Sec.~\ref{sec:exp} describes the current \neb reactor experiments, relying on the flux predictions presented in Sec.~\ref{sec:flux}; the observation of the energy distortion is reported in Sec.~\ref{sec:struct}, followed by a critical analysis of its possible origin in Sec.~\ref{sec:origin}; the reactor flux models are revisited in Sec.~\ref{sec:models} as the most likely cause; Finally, Sec.~\ref{sec:summary} summarizes the state-of-the-art and discusses about the experimental data required to gain further knowledge on the \neb reactor flux and the origin of the energy structure.

\section{A new generation of reactor neutrino experiments}
\label{sec:exp}

\begin{figure}[htbp]
\begin{center}
\includegraphics[width=60mm]{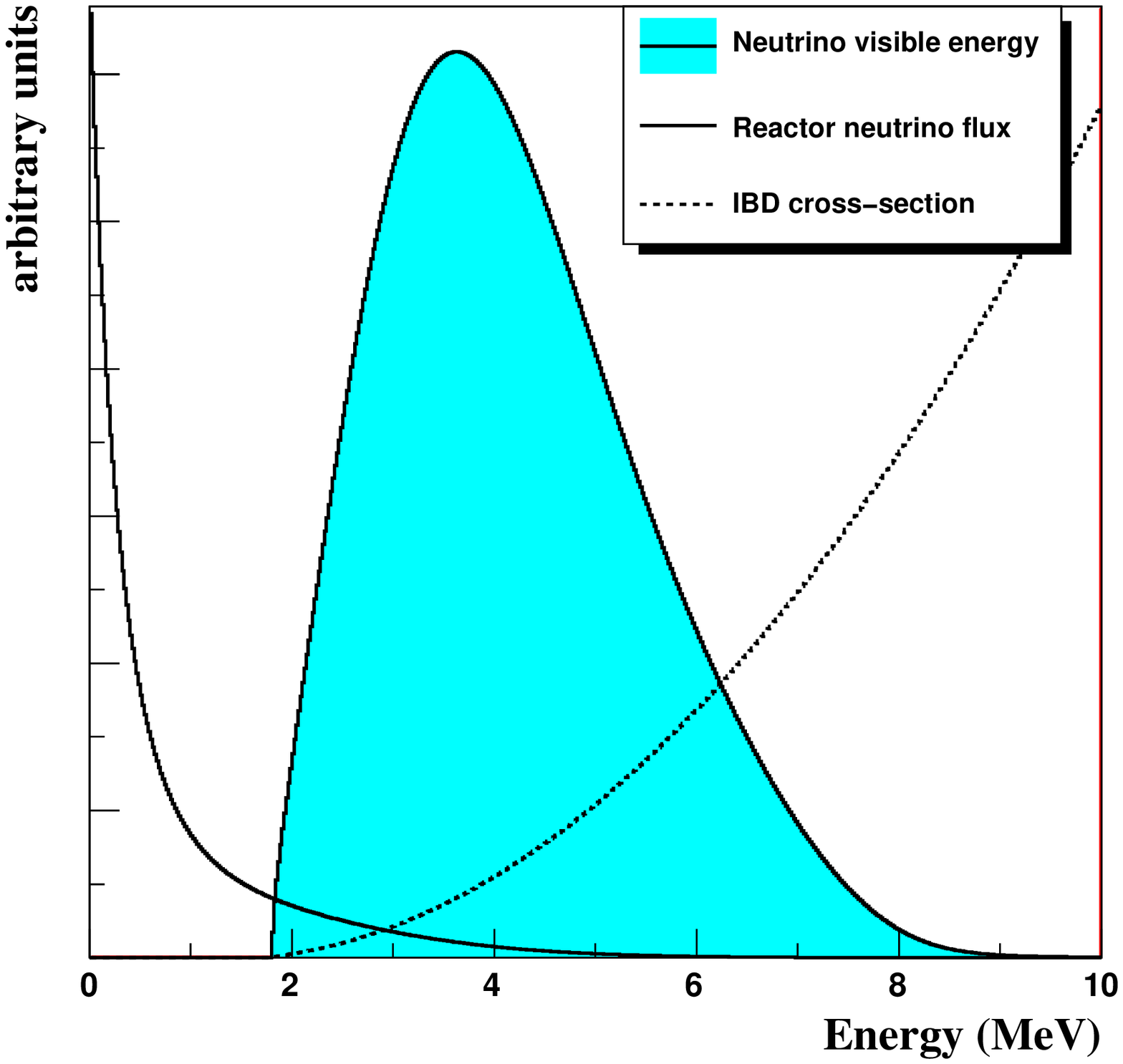}
\includegraphics[width=60mm]{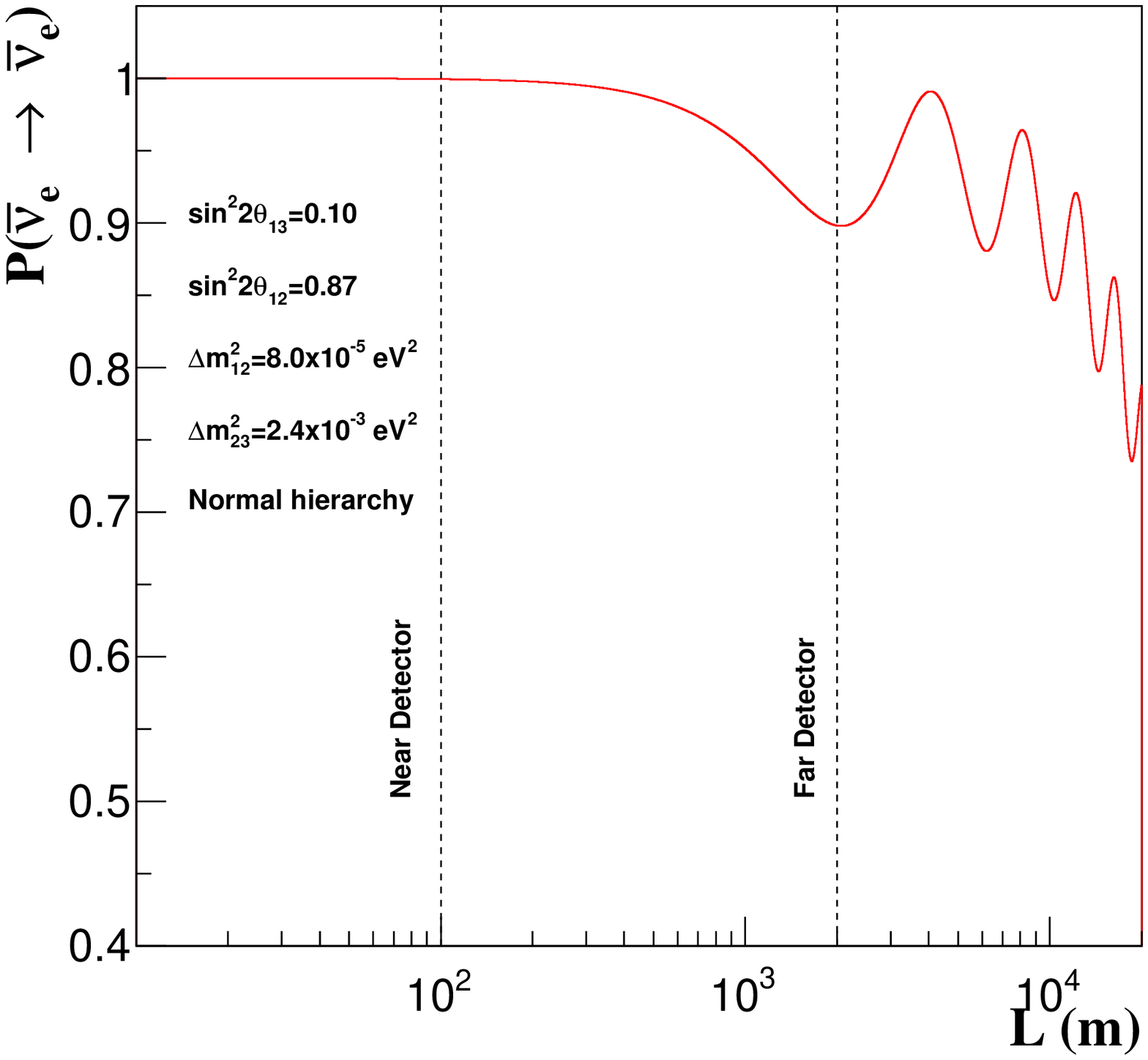}
\end{center}
\caption{\label{fig:nuexp} Left: $\bar\nu_e$ visible spectrum as a result of the flux shape and IBD cross-section. Right: $\bar\nu_e$ survival probability for $E_{\bar\nu}=4$ MeV, as a function of the distance $L$ for an arbitrary value of $\sin^22\theta_{13}=0.10$. This probability assumes a three flavor neutrino scenario and normal hierarchy of the neutrino masses.}
\end{figure}

The most common way of detecting reactor neutrinos is via the inverse beta decay (IBD): $\bar\nu_e +p \rightarrow n + e^+$. When this reaction takes place in liquid scintillator doped with $\sim$1\% of Gadolinium, it produces two signals separated by about $\sim30$ $\mu$s: the first one due to the $e^+$ and its annihilation (prompt signal), and the second one due to the $n$ capture in a Gd nucleus (delayed signal). This characteristic signature yields a very efficient background rejection. The prompt energy deposition ($E_{e}$) relates directly to the interacting antineutrino energy ($E_{\bar\nu}$):  $E_{e}\simeq E_{\bar\nu}+(M_p-M_n-M_e)+2M_e$, where $M_p$, $M_n$ and $M_e$ are the proton, neutron and electron masses, respectively. The observed energy spectrum is the convolution of the reactor \neb flux and the IBD cross section. As shown in the left panel of Fig.~\ref{fig:nuexp}, the mean energy of the $\bar\nu_e$ spectrum is around $4$ MeV, which corresponds to a prompt energy $E_{e}$ of $\sim$3 MeV. For this energy, the oscillation effect due to $\theta_{13}$ starts arising at $L\sim0.5$ km and reaches the first maximum around 2 km, where the effect of $\theta_{12}$ is still negligible as can be seen in the right panel of Fig.~\ref{fig:nuexp}. Therefore, neutrino reactor experiments with short baselines offer a clean laboratory to search for $\theta_{13}$.     

In spite of its characteristic signature, the IBD signal can be mimicked by the accidental and correlated backgrounds. The accidental background stands for the random coincidence of a positron-like signal coming from natural radioactivity, and the capture in the detector of a neutron created by cosmic muon spallation in the surrounding materials. The correlated background consists of events which may mimic both the prompt and the delayed signals of the IBD. Along with the stopping muons, the fast neutrons and cosmogenic isotopes, both generated in muon interactions, are the main sources of this background. Fast neutrons entering the detector lead to proton recoils, thus faking a prompt signal, before being captured. Muons crossing the detector can produce long-lived $\beta$-n decay isotopes, like $^9$Li and $^8$He. As the half-life of such cosmogenic isotopes is $\sim$100 ms, their decay cannot be associated to the muon interaction. 


The sensitivity to the $\theta_{13}$-driven oscillation is optimized by detecting a deficit in the expected neutrino events around 2 km away from the nuclear power plant (\emph{far} detector), as shown in right panel of Fig.~\ref{fig:nuexp}. However, some of the largest systematic errors in reactor experiments arise from the uncertainties in the original $\bar\nu_e$ fluxes. In order to reduce them, a relative comparison between two or more identical detectors located at different distances from the reactors becomes critical. As originally proposed in \cite{kr2det}, a \emph{near} detector placed a few hundred meters away can measure the fluxes before any oscillation takes place. The comparison between the far and near detectors leads to a breakthrough in the sensitivity to $\theta_{13}$, as all the fully correlated systematic uncertainties cancel out. Further steps in the sensitivity optimization rely on reducing the relative detection efficiency and energy scale uncertainties of the detectors, as well as on minimizing the backgrounds.  


Following the above ideas, a new generation of reactor experiments are running since 2010. In China, the Daya Bay experiment \cite{db} has built a far site and two near sites meant to measure the $\bar\nu_e$ fluxes from the 6 cores (17.4 GW$_{th}$ in total) of the three power plants existing in the area. The Double Chooz experiment \cite{dcIII} operates two identical detectors located 400 m (since 2014) and 1050 m away from the two 4.25 GW$_{th}$ reactor cores of the CHOOZ nuclear plant in France. RENO \cite{reno} also consists of two identical detectors measuring the antineutrino fluxes generated at the 6 cores (17.3 GW$_{th}$ in total) of the Youngwang nuclear plant in South Korea. Although there are some differences in the detector designs of the tree experiments, they all rely on the same principles and technology. The detectors are divided into three concentric volumes: the target (the inner-most volume filled with Gd-doped liquid scintillator), the $\gamma$-catcher (filled with undoped liquid scintillator) and the buffer (filled with mineral oil). The light produced by interactions in the liquid scintillator is read out by a number of photomultiplier tubes (PMTs) located in the buffer walls.  

\section{Antineutrino flux prediction}
\label{sec:flux}

In a nuclear reactor, about 6 antineutrinos from $\beta$-decays are generated per fission, releasing an average energy of about 200 MeV. As the unstable fission products are rich in neutrons, they undergo $\beta^{-}$ decays generating a nearly pure electron antineutrino flux. Only four isotopes, whose fission products can produce \neb with energies above the IBD threshold (1.8 MeV), contribute to more than 99\% of the flux: \Urfive, \Punine, \Ureight ~and \Puone. However, such a flux consists of a superposition of thousands of $\beta$-decay branches. A fraction of the neutrons produced in the \Urfive ~fissions is captured by \Ureight, giving place to mostly \Punine. Thus, the core burns \Urfive ~while accumulating \Punine ~as it is operated, in the so-called burnup process. Apart from these two main isotopes which make up about 90\% of the flux, the \Puone ~and \Ureight ~fissions contribute to the remaining 10\%. From a practical point of view, this implies that an accurate reactor flux prediction relies on two main aspects: 1) the simulation of the time evolution of the core fuel composition (i.e, the contributions of each one of the four main isotopes), and 2) the knowledge of the $\beta$ spectra associated to the decay chains of the fission products.   

In order to compute the expected neutrino flux in a reactor experiment like Daya Bay, Double Chooz or RENO, three main ingredients need to be taken into account: 1) the detector-related normalization terms, 2) the reactor flux as a function of time, and 3) the IBD cross section. In absence of oscillations, the number of expected antineutrinos from a nuclear core can be described as:

\begin{equation}
\label{eq:prediction}
N^{exp} = 
	\frac{\epsilon N_{p} }{4\pi} 
	 \frac{1}{L^2} \frac{P_{th}}{\langle E_{f} \rangle } 
\times \langle \sigma_f \rangle
\end{equation}

\noindent where $\epsilon$ is the detection efficiency,  $N_{p}$ is the number of protons in the target, $L$ is the distance to the center of the reactor, and $P_{th}$ is the thermal power. $\langle E_{f} \rangle$ is the mean energy released per fission:

\begin{equation}
\label{eq:mpfiss}
\langle E_{f} \rangle = \sum_{k} \alpha_k  \langle E_{f} \rangle_k,
\end{equation}

\noindent where $\alpha_{k}$ is the fractional fission rate of the $k^{th}$ 
isotope ($k=$ \Urfive, \Punine, \Ureight, \Puone). The mean cross section per fission $\langle \sigma_f \rangle_k$  is defined as:  

\begin{equation}
\label{eq:crossperfiss}
\langle \sigma_f \rangle = \sum_k \alpha_k \langle \sigma_f \rangle_k 
=  \sum_k \alpha_k \int_{0}^{\infty} dE \, S_{k}(E) \, 
\sigma_{I\!B\!D}(E),
\end{equation}
\noindent where $S_{k}(E)$ is the 
reference spectrum of the $k^{th}$ isotope and $\sigma_{I\!B\!D}$ is the 
inverse beta decay cross section. The three variables $P_{th}$, $\langle E_{f} \rangle$ and $\langle \sigma_f \rangle$ are time dependent, with  $\langle E_{f}  \rangle$ and $\langle \sigma_f \rangle$ depending on the evolution of the fuel composition in the reactor and $P_{th}$ depending on the operation of the reactor.


The current reactor experiments have used the reference spectra $S_{k}(E)$ from \cite{Mueller,Huber} as an input to their oscillation analyses. However, as far as the determination of these spectra is concerned, $S_{k}(E)$ can be expressed as the sum of the contributions from all the fission products ($N_f$):

\begin{equation}\label{eq:Sk}
S_{k}(E)=\sum_{f=1}^{N_{f}}\mathcal{A}_{f}\times S_{f}(E),
\end{equation}

\noindent where $\mathcal{A}_{f}$ is the activity of the fission product and the spectrum $S_{f}(E)$ of each fission product is in turn a sum of $N_b$ $\beta$-branches connecting the ground state (or an isomeric state) of the parent nucleus to different excited levels of the daughter nucleus:

\begin{equation}
\label{eq:Sfp}
S_{f}(E)=\sum_{b=1}^{N_b} BR_{f}^b\times S_{f}^b(Z_{f},A_{f},E_{0f}^b,E),
\end{equation}

\noindent being $BR_{f}^b$ and $E_{0f}^b$ the branching ratio and the endpoint energy of the $b$ branch of the $f$ fission product, respectively, and $Z_{f}$ and $A_{f}$ the charge and atomic number of the parent nucleus. It is worth noticing that Equations (\ref{eq:Sk}) and (\ref{eq:Sfp}) are valid for both electron and antineutrino spectra. The beta decay spectrum $S_{f}^b$ for a single transition in a nucleus with endpoint energy $E_{0f}^b=E_e - E_{\nu}$ is then:

\begin{equation}
S_f^{b}(E_e,Z_f,A_f) =  S_0 (E_e) F(E_e,Z_f,A_f) C(E_e) (1+\delta(E_e,Z_f,A_f)), 
\label{allowed}
\end{equation}
where $S_0$ is a normalization constant taking into account the phase space \cite{Mueller, Hayes14}, $F(E_e,Z_f,A_f)$ is the Fermi function accounting for the Coulomb interaction of the outgoing electron with the charge of the daughter nucleus, and $C(E_e)$ is a shape factor \cite{schopper} for forbidden transitions due to additional lepton momentum terms ($C(E)=1$ for allowed transitions). Beyond these terms, some additional effects need to be considered for precision studies: this is the role of the $\delta(E_e,Z,A)$ factor. It accounts for the radiative (R), finite size (FS) and weak magnetism (WM) corrections: $\delta(E_e,Z_f,A_f)=\delta_{\mathrm{R}}+\delta_{\mathrm{FS}} +\delta_{\mathrm{WM}}$. The R corrections are due to the emission of virtual and real photons by the charged particles present in the $\beta$-decay, and it is computed in \cite{sirlin}. The FS correction accounts for the finite size of the nucleus, as the electric charge and hypercharge are not point-like \cite{vogel84,wilkinson}. The WM term refers to the induced current yielding the largest contribution to the shape of the $\beta$ spectrum \cite{vogel84,holstein}.


Finally, the simplified form from Vogel and Beacom \cite{Vogel1999} can be used to describe the IBD cross section:
\begin{equation}
\sigma_{I\!B\!D}(E_{\nu}^{true}) =   E_{e^{+}}K \sqrt{E_{e^{+}}^2 - 
m_{e}^2} ,
\end{equation}
\noindent where
\begin{equation}
E_{e^{+}}  = \frac{1}{2} \left( \sqrt{m_{n}^2 - 4 m_{p} \left( -
E_{\nu} + \Delta + \frac{\Delta^2 - m_e^2}{2m_p} \right)}-
m_n \right)
\end{equation}
\noindent and $m_e$ and $E_{e^+}$ are the positron mass and energy. The variables $m_n$ and $m_p$ are the masses of the neutron and proton with $\Delta=m_n - m_p$. The constant $K$ is inversely proportional to the neutron lifetime. Using the MAMBO-II measurement of the neutron lifetime \cite{MAMBOII} leads to $K=0.961\times10^{-43}$ cm$^{2}$ MeV$^{-2}$.

\subsection{Reactor flux models}

In order to predict the reference spectra $S_{k}(E)$, two different approaches are developed. The \emph{ab initio} or summation method takes advantage of the available information on the $\beta$ decays of each fission fragment (nuclear databases), summing over each nuclide's individual spectrum to obtain an aggregate spectra. On the other hand, the so-called conversion method exploits the aggregate $\beta$ spectra measured in the Institut Laue-Langevin (ILL) \cite{SchreckU5-1, SchreckU5-2, SchreckU5Pu9, SchreckPu9Pu1}, fitting the data to a set of virtual $\beta$ branches and converting the result into the corresponding antineutrino spectra (e.g.~\cite{Conv}). While it is worth noticing that both methods rely on measured $\beta$ spectra, the conversion approach yields the most precise results since the uncertainties are constrained by the ILL measurements. Although this is reviewed in this work, the errors associated to the conversion method have been claimed to be at the level of 2-3\%. As the nuclear databases are known to suffer from a lack of relevant data (concerning both $\beta$ decays and fission yields) and from the need of more precise measurements, the summation method provides typically an envelope error of about 10-20\%. Recently, there have been improvements in both the conversion and summation techniques \cite{Mueller,Fallot}, being one of the main goals to optimize the results from the current reactor experiments.  

Given the limitations of the \emph{ab initio} approach, the reactor antineutrino spectra have been estimated historically relying on the total electron spectra associated with the beta decays of all fission products of \Urfive, \Punine, and \Puone. Such $\beta$-spectra were obtained at ILL by irradiating thin target foils of these isotopes with thermal neutrons. As \Ureight ~nuclei undergo fission with fast neutrons, the associated spectrum could not be measured at that time and therefore its prediction has been typically based on the summation method. Same applies to all spectra above 8 MeV, as the ILL measurements were performed only up to this energy. In \cite{Mueller}, a mixed approach has been developed combining the precise reference of the electron spectra from ILL with the physical distribution of beta branches of all fission products provided by the nuclear databases. This new analysis has provided a better handle on the systematic errors of the conversion, and a new set of reference spectra for \Urfive, \Punine, \Puone ~and \Ureight ~(although the latter is still based on a purely summation technique). While the shapes of the spectra and their uncertainties are found to be comparable to that of the previous analysis of the ILL data, the normalization is shifted by about +3\% on average, thus leading to the reactor neutrino anomaly. The re-evaluation of short-baseline reactor data in the light of these new reference spectra reveals a deficit in the number of observed antineutrinos, which might be explained in terms of sterile neutrino oscillations. One of the main reasons for this normalization shift is the treatment of the corrections $\delta(E_e,Z_p,A_p)$ in Eq.~\ref{allowed}, and in particular the WM term. These corrections have been further investigated in \cite{Huber}, deriving a consistent set of reference spectra in both shape and normalization. To complete the picture of the state-of-the-art conversion method, it must be noticed that the cumulative $\beta$ spectrum of the fission products of \Ureight ~has been finally measured in \cite{Haag}, in the range from 2.875 MeV to 7.625 MeV.


Despite the larger uncertainties, the summation method is still a powerful tool to predict reactor antineutrino fluxes. To start with, this approach provides estimations of the reference spectra which are independent from the measurements at ILL. As these measurements are unique, a cross check based on nuclear databases is specially valuable. The summation method is also the only way to predict the antineutrino energy spectra beyond 8.0 MeV for \Urfive, \Punine ~and \Puone, and beyond  7.6 MeV for \Ureight. While it is true that some authors (e.g. \cite{Mueller,Huber}) have provided polynomial parameterizations of the spectra that can be used to extrapolate the predictions above 8 MeV (as typically done by reactor experiments), such an extrapolation is not physically motivated and the associated error cannot be estimated in a robust way. Furthermore, the binning of 250 keV in which the ILL spectra were originally published is large enough to hide possible structures coming from some contributions to the reactor fluxes. Taking into account all these considerations, the summation technique has become a main tool to shed light on the reactor antineutrino spectra and the nuclear databases have been recently improved by new $\beta$-decay measurements. In particular, a new set of reference energy spectra has been obtained in \cite{Fallot} taking into account the new measurements of the $^{102;104;105;106;107 }$Tc, $^{105}$Mo and $^{101}$Nb nuclei. These measurements are taken with the Total Absorption Technique (TAS), insensitive to the Pandemonium effect \cite{pande} which typically affects the $\gamma$ spectrometry with Ge detectors. Beyond the relevant improvement in the summation-based reference spectra, the work in \cite{Fallot} highlights the need of new TAS measurements.   

\begin{figure} 
  \begin{center}
    \includegraphics[width=8cm]{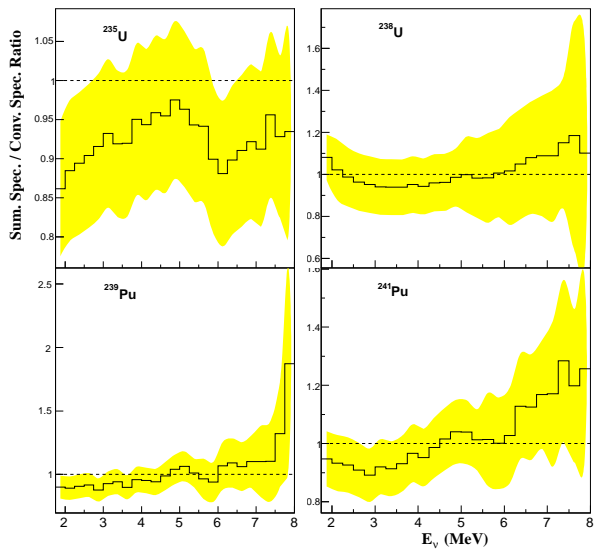}
    \caption{ Ratio of summation-derived spectra (from \cite{Fallot}) to the start-of-art conversion-derived spectra (from \cite{Huber} and \cite{Haag}), as a function of the \neb energy. The shadowed band shows the 1$\sigma$ error.}
    \label{fig:RefRatio}
  \end{center}
\end{figure}

In order to compare the state-of-art reference spectra, the ratio of the summation spectra derived in \cite{Fallot} to the conversion-based predictions (\Urfive, \Punine ~and \Puone ~from \cite{Huber} and \Ureight ~from \cite{Haag}) is shown in Fig.~\ref{fig:RefRatio}. Both set of predictions agree within 1$\sigma$, although no strong conclusions can be settle given the large error bands.

\section{Observation of an energy structure around 5 MeV}
\label{sec:struct}

In the Neutrino 2012 conference, the RENO collaboration mentioned the observation of an excess in the number of antineutrinos between 4 and 6 MeV of the positron energy spectrum \cite{RENONu12}, with respect the to flux prediction. In Neutrino 2014, both RENO and Double Chooz collaborations reported and quantified such an excess \cite{RENONu14,DCNu14}. Daya Bay also presented a similar energy structure at the ICHEP 2014 and at NuTel 2015 \cite{DBNuTel15} conferences. 

The first journal publication of this effect has been provided by Double Chooz in \cite{dcIII}. Using the data from the far detector, a best fit value of the flux normalization of 9$\pm$2\% (with respect to the central value prediction) is quoted between 4.25 and 6 MeV. This translates into a significance of 3$\sigma$. The energy distortion is consistent with the ones presented in previous Double Chooz publications \cite{dcI, dcII,dcH}, where a significant measurement of the data-prediction discrepancy was not possible due to the limited statistics and the non-optimized detector energy response. Daya Bay has also released a paper \cite{DBFlux} on the reactor neutrino flux measurement, concerning both the normalization and the energy spectrum. The measured prompt energy spectrum in the near detectors shows a deviation from the reactor models with a significance beyond 2$\sigma$ over the full energy range, and around 4$\sigma$ between 4 and 6 MeV. The excess in this energy range has been estimated to be about 1\% of all events in both the near and far detectors. Once corrected by the $\theta_{13}$ oscillation effect, the energy spectra measured in the near and far detectors are consistent. Two reactor predictions have been considered in Daya Bay, one based on the conventional ILL models and another one based on the re-evaluations provided in \cite{Mueller,Huber}. The disagreement between the data and the prediction arises in both cases, being the significance of the deviations very similar. The significance has also been computed adopting two different approaches, one relying on the contribution of the $\chi^{2}$ of each energy bin and another one on the p-values within local energy windows, yielding consistent results. In addition, the RENO collaboration has shown in the the proceedings of Neutrino 2014 \cite{RENONu14} the energy structure in the 4-6 MeV energy range for both the near and far detectors. The observed excess of \neb is consistent among the two, and the significance of the excess at the near detector is estimated to be 3.5$\sigma$. The excess of \neb with respect to the total expected flux is quoted as 2.3$\pm$0.4 (data) $\pm$ 0.5 (prediction)\% for the near detector, and 1.8$\pm$0.7 (data) $\pm$ 0.5 (prediction)\% for the far detector.

\begin{figure} 
  \begin{center}
    \includegraphics[width=6cm]{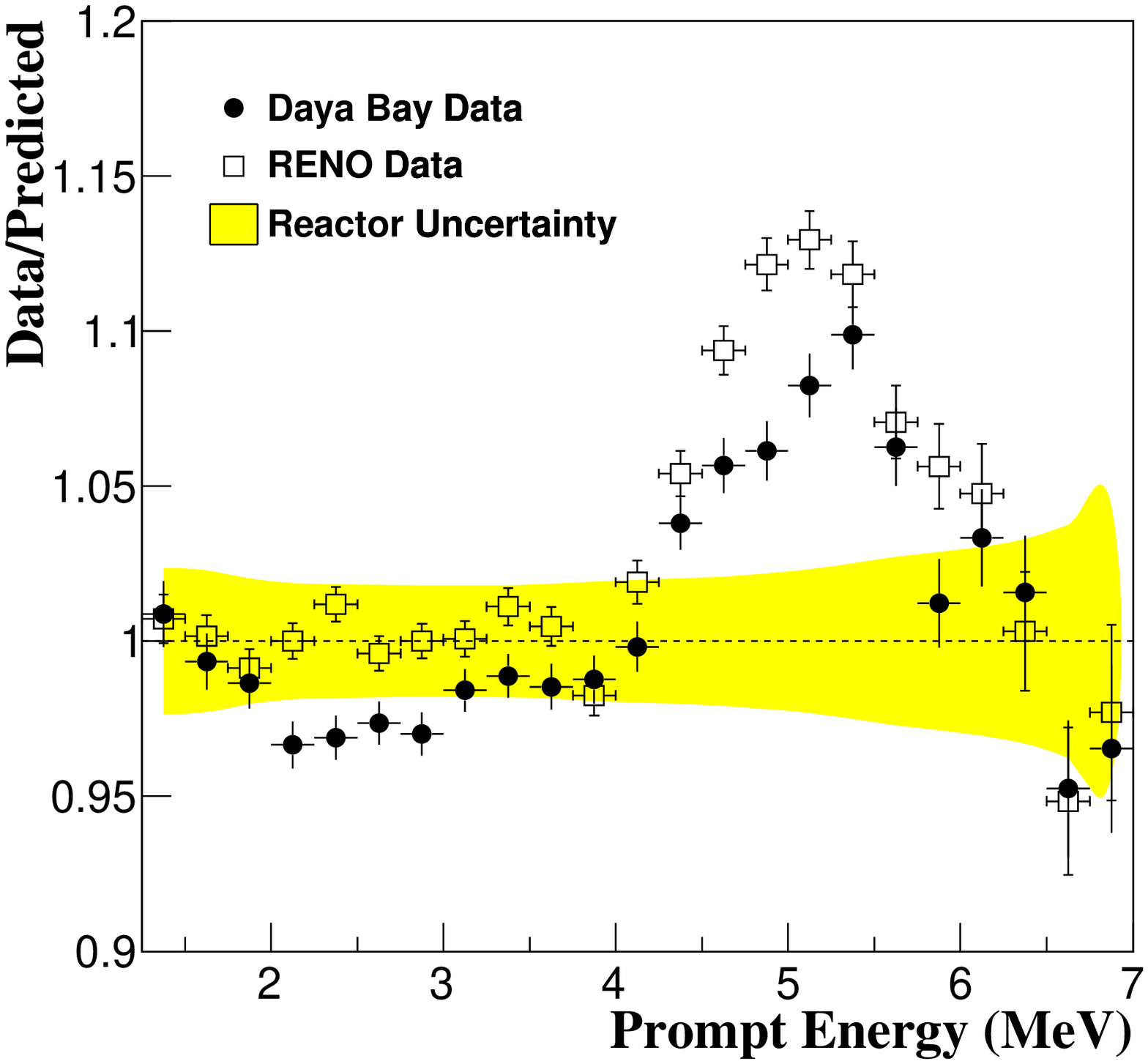}
    \includegraphics[width=6cm]{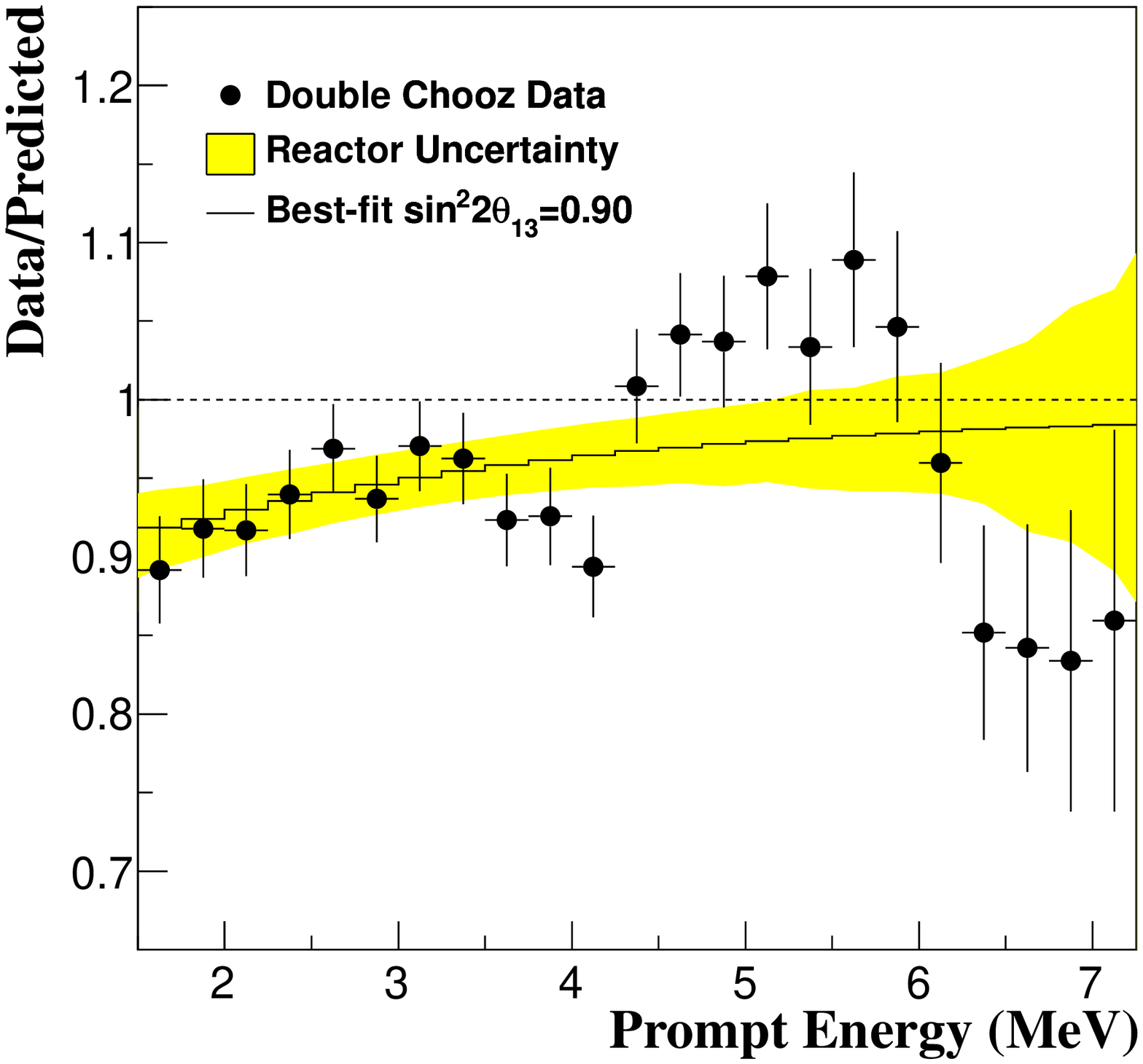}
    \caption{ Ratio of background-subtracted \neb candidates to non-oscillation prediction, as a function of the IBD prompt energy. Left: Daya Bay and RENO ratios for \neb candidates observed at the near sites. Right: Double Chooz ratio for \neb candidates measured at the far detector. The shadowed region represents the typical reactor error derived from the \cite{Huber} and \cite{Haag} reference \neb spectra, which is dominant for the considered energy region.}
    \label{fig:SpectrumRatio}
  \end{center}
\end{figure}

The ratio of the background-subtracted \neb candidates spectrum to the non-oscillation prediction is shown in Fig.~\ref{fig:SpectrumRatio} for the Daya Bay and RENO near detectors and for the Double Chooz far detector. The excess reported by the three collaborations amounts to about 10\% over the expected number of \neb in this energy range, and both RENO and Daya Bay have observed consistent structures in their near and far detectors. It might be argued that the excess observed in RENO is larger than in Daya Bay and Double Chooz. However, given the discrepancy between RENO and Daya Bay in the 2-4 MeV range (see Fig.~\ref{fig:SpectrumRatio}), such a difference in the amplitude of the distortion might be due to differences in the flux predictions beyond the 4-6 MeV window. In order to compare the observed distortions in a robust and quantitative way, the \neb prediction of the three reactor experiments should be based on the same reactor model (which comprises not only the reference spectra $S_k(E)$, but also the simulation of the reactor core evolution, the treatment of the spent fuel, etc). It is also worth noticing that the significances of the excess quoted by the three experiments cannot be directly compared, as they are computed in different ways. Daya Bay normalizes the predicted spectrum to the observed number of events, thus evaluating the discrepancy in terms of the energy spectrum between 4 and 6 MeV, and not the total rate. On the other hand, Double Chooz performs an evaluation based on the total predicted and expected rates in the 4.25-6.00 MeV energy window. Independently of how the distortion significance is estimated, it is limited by the uncertainties in the flux prediction, which are at the level of 2-3\% for both the rate and the spectral shape.

 As the three experiments detect the reactor \neb in the same way and with very similar detectors, the possible causes of this energy structure are common and might include, in principle, detector and/or background issues. An explanation in terms of the reactor flux prediction (incompleteness of the model or underestimation of the uncertainties) would be also correlated among the three experiments, as they all rely on the conversion method to obtain the \Urfive, \Punine ~and \Puone ~antineutrino spectra. There are however two differences. Firstly, Double Chooz uses in \cite{dcIII} the measurement in \cite{Haag} to derive the \Ureight ~\neb spectrum, while Daya Bay and RENO take the summation-based spectrum from \cite{Mueller}. Secondly, Double Chooz constrains the flux normalization to the measurement in Bugey4~\cite{Bugey4}, taken 15\,m away from the core. This is why Double Chooz quotes a flux normalization error of 1.7\%, thus reducing the errors in Daya Bay and RENO (2.7\% and 2.0\%, respectively). Beyond this, Daya Bay has also explored some variations to the reference model but concluded that the structure still remains. In particular, the local distortion around 5 MeV cannot be described extending the reactor model with a single $\beta$-branch or a mono-energetic line.


While this disagreement between data and flux models needs to be investigated, it must be noticed that the impact on the $\theta_{13}$ mixing angle is negligible. As demonstrated in \cite{dcIII} for Double Chooz, even with only the far detector being used for the oscillation analysis, the $\theta_{13}$ measurement is not affected by the energy distortion. This can be easily understood since the amplitude of the $\theta_{13}$-driven oscillation is vanishing around 5 MeV. In the case of Daya Bay and RENO, whose analyses involve both near and far detectors, the impact of the energy structure is even smaller as the role of the flux prediction is not as relevant due to the inter-detector comparison.

\section{Possible sources of the energy structure}
\label{sec:origin}

The spectral shape of the energy structure at 5 MeV cannot be produced by
any standard neutrino oscillations scenario, even considering sterile neutrinos. In particular, it has been observed by Daya Bay and RENO at two different baselines (the near and far detectors). Therefore, it can be assumed that the discrepancy between the data and the flux models might be due to one of these reasons: 1) the existence of non-standard IBD interactions, 2) a detection issue distorting the energy scale, 3) an unaccounted background source, and 4) missing contributions to the reactor models. Being the reactor \neb spectrum uncertainty of the order of 2-3\% and the maximum deviation between data and prediction around 10\%, the current reactor experiments cannot establish this discrepancy beyond a significance of $\sim$4$\sigma$. However, the available data allow for a dedicated analysis on the possibles causes of the discrepancy. The current reactor experiments have been capable of reinforcing the case for a reactor-model explanation, while disfavoring other possible causes, namely the misinterpretation of the detector response and the incompleteness of the background model. The dedicated studies addressing the possible sources of the prompt energy spectrum distortion are described below.

\subsection{Antineutrino interactions}

An unaccounted or non-standard neutrino interaction in the detectors of the reactor experiments might lead to an excess of observed neutrinos. In the energy range of reactor \neb (below 10 MeV), the typical cross section of charged and neutral current neutrino interactions follow an increasing pattern with energy after a given threshold. This kind of trend can hardly explain a bump-like excess around 5 MeV in the positron energy spectrum. Within the target volume of the detectors, the antineutrinos can interact basically with H, C and Gd. In the $\gamma$-catcher, only with H and C. As Double Chooz has observed the energy structure using neutrons captures in Gd and in H \cite{dcH}, an unaccounted interaction with Gd can be excluded. The antineutrinos might interact with some C isotope with enough energy to separate one neutron, remaining the final nuclei in excited state. The combination of the de-excitation $\gamma$ and the neutron might mimic the IBD signal. However, the rate of such process should be rather small (as there are not empirical evidences in the current experiments) and could not explain the $\sim$10\% excess.

\subsection{Energy scale}

A detector-related issue affecting the energy scale might also explain, in principle, the energy structure. However, some studies performed by Daya Bay and Double Chooz rule out this possibility. In \cite{dcIII}, the accuracy of the energy scale around 5\,MeV has been confirmed by spallation neutrons captured on carbon, which  occur predominantly in the $\gamma$-catcher volume as the capture cross section is smaller than on Gd. The C captures result in an energy peak at 5\,MeV, whose agreement between data and MC simulation has been found to be within 0.5\,\%. Along the same lines, Daya Bay, Double Chooz and RENO have also checked the energy reconstruction by means of the $\beta$ decays of $^{12}$B collected in data, showing no energy distortion when compared to the corresponding simulation. This is consistent with the fact that any nonlinear effect, due to the scintillator properties or the electronics response, is observed for energies above 4 MeV. Furthermore, the energy resolution estimated with the collected data is also in good agreement with that of the Monte Carlo.

\subsection{Background model}

The events found in the 4-6 MeV range fulfill all the IBD characteristics, in particular concerning the neutron capture time and distance distribution, and the spatial distribution of the prompt signals. Thus, the three collaborations disfavor the hypothesis of an unaccounted background contribution. In addition, the reactor off data taken in Double Chooz allows for an independent and inclusive background measurement, thus accounting even for possible unknown sources \cite{dc2off}. The measured total rate in \cite{dcIII} according to the candidates selection cuts is computed to be 0.75$\pm$0.37 events/day. While keeping the independence with the background model, this background measurement is slightly modified by means of a Reactor Rate Modulation (RRM) \cite{dcrrm} fit: 0.90$^{+0.43}_{-0.36}$ events/day. These total background measurements are lower than the sum of the individual background sources accounted for in the background model (accidental coincidences, fast-neutrons/stopping-muons, and cosmogenic isotopes): 1.6$^{+0.41}_{-0.17}$ events/day (1.7$\sigma$ discrepancy with respect to the reactor off measurement). Therefore, the existence of an unaccounted background source, leading to the excess around 5 MeV, is strongly disfavored. Beyond this comparison with the inclusive background measurement, dedicated studies with reactor on and reactor off data have been developed to look for direct indications of unknown background sources. No significant evidences have been found. 

\subsection{Reactor flux model}

\begin{figure} 
  \begin{center}
    \includegraphics[width=6cm]{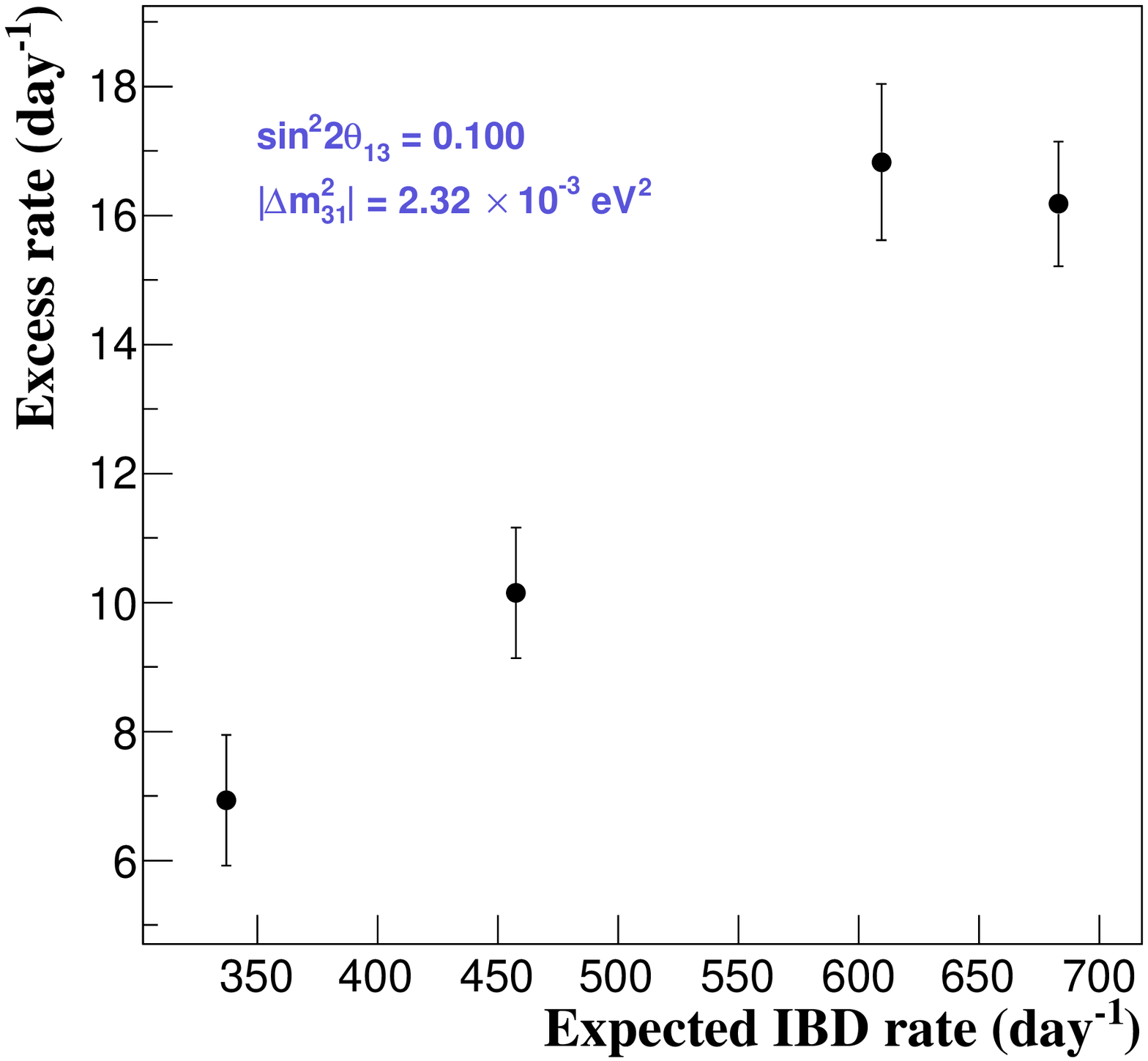}
    \includegraphics[width=6cm]{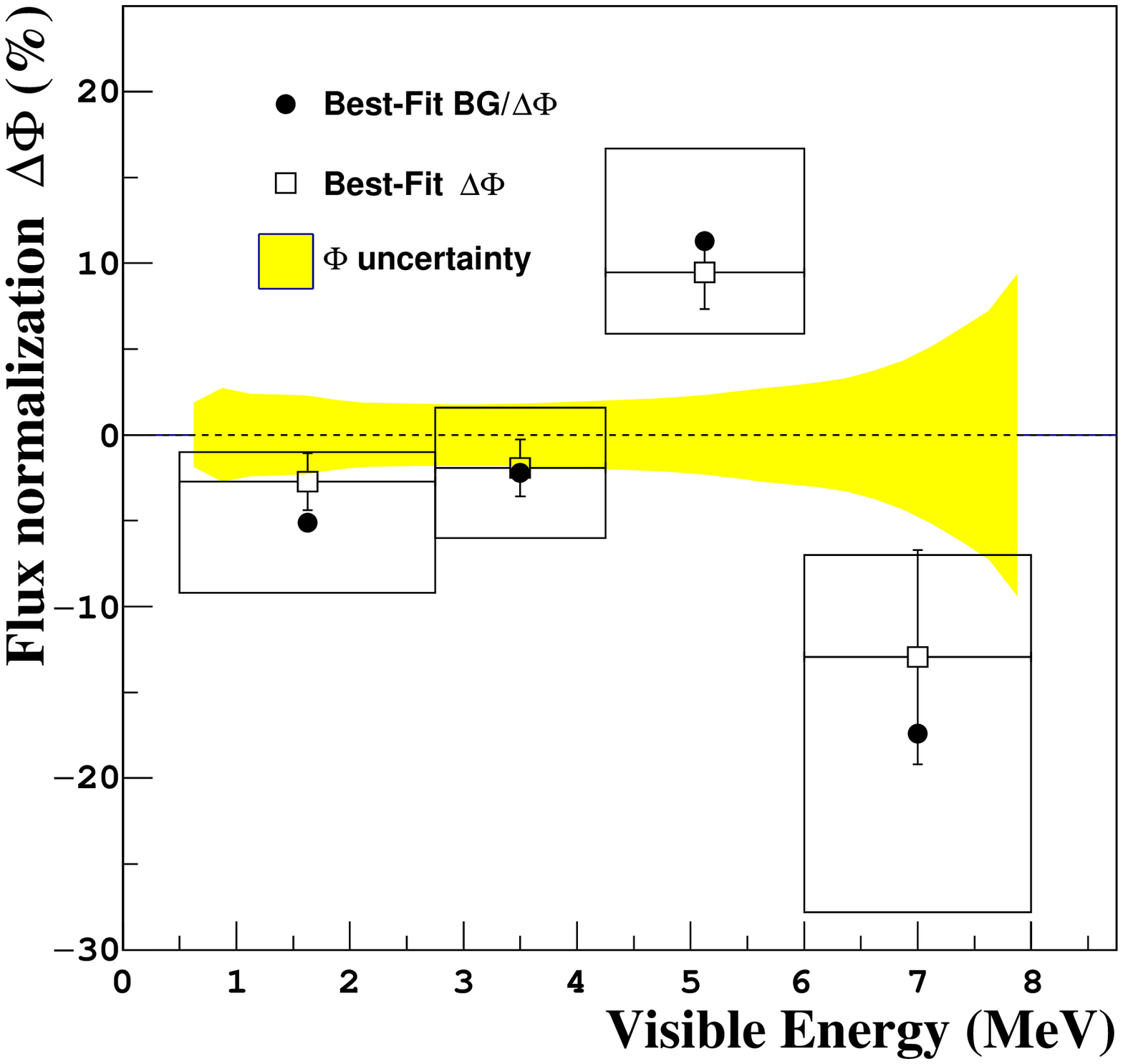}
    \caption{ Correlation of the \neb excess around 5 MeV with the reactor power. Left: \neb excess in the RENO near detector as a function of the expected IBD (data from \cite{RENONu14}). Right: RRM best fit values of the reactor flux normalization (with respect to the central value prediction) in the far detector of Double Chooz (data from \cite{dcIII}). Results with and without the background (BG) model constraint are shown with empty squares and solid dots, respectively.}
    \label{fig:reaccorr}
  \end{center}
\end{figure}

The remaining possible cause of the energy distortion is that one of an additional reactor $\bar\nu_{e}$ component beyond the current model. In particular, if the excess around 5 MeV is due to an unaccounted reactor contribution, it must be correlated to the reactor power. On the other hand, if it is due to an unknown background, the rate of the excess should be independent of the power. Such a correlation has been demonstrated by Daya Baya, Double Chooz and RENO, by estimating the excess for different reactor powers. Daya Bay has shown in ICHEP 2015 the time stability of the prompt energy spectrum and the time distribution of events for two different energy windows (4.5-5.5 MeV and 3.0-4.0 MeV), proving that the structure remains the same over time, and thus for different conditions of the reactors operation. In \cite{dcIII}, the Double Chooz collaboration shows the correlation of the excess with the reactor power in a flux-model independent way, by parameterizing the spectrum and measuring an effective excess for different reactor conditions. Consistent results are found for \neb candidates obtained with neutron captures in Gd and H. RENO has also reported in Neutrino 2015 and NuTel 2015 such a correlation by means of the measurement of the excess for different reactor powers, as shown in left panel of Fig.~\ref{fig:reaccorr}.       

As the Double Chooz RRM analysis utilizes the correlation between the observed rate and the thermal power to derive both the mixing angle $\theta_{13}$ and the total background rate, it can be used to test the hypothesis of a bias in the flux prediction. In particular, it can confront the data to the background model and the flux model at the same time, thus providing indications about the most likely cause of the energy structure. In \cite{dcIII}, five independent RRM fits have been carried out in different energy regions, constraining $\sin^22\theta_{13}$ to the best fit value in \cite{db} while leaving as free parameters both the total background rate and a flux normalization term (with respect to the central value of the flux model). The best fit values of the background rate are fully consistent with both the background model and the reactor off measurement, while the best fit values for the flux normalization deviates (2$\sigma$) from the prediction in the 4.25-6.00 MeV window, as shown in the right panel of Fig.~\ref{fig:reaccorr}. This result is consistent with the reported correlation between the excess and the thermal power, thus reinforcing the case for a flux model bias and disfavoring again the background model as the source of the energy distortion. If one constrains the total background rate to the background model, the discrepancy between the flux model and the RRM best fit value is increased to 3.0$\sigma$. 

\section{Reviewing the reactor flux predictions}
\label{sec:models}

As discussed above, the most likely explanation for the energy structure is that one of unaccounted contributions in the reactor flux models. Hereafter, this work assumes this is the unique source of the data-prediction disagreement around 5 MeV. Under this well motivated assumption, the different approaches to the flux estimations need to be reviewed from a critical point of view. To start with, the errors of the conversion-based reference spectra must be re-evaluated somehow, as the current quoted uncertainties do not cover the observed energy distortion. This is obviously related to the identification of the possible missing pieces in the reactor models. Such contributions in the conversion-based flux might be related to: 1) the aggregate $\beta$ measurements, 2) the conversion procedure itself, and 3) the nuclear corrections, mostly related to the forbidden decays. To shed light on all these possibilities, the summation-based predictions might play a major role, but the associated limitations need to be taken into account. All these aspects are discussed below.     

\subsection{The aggregate spectra}

As the  \Urfive, \Punine ~and \Puone ~reference spectra obtained by means of the conversion method rely on the ILL data, any issue affecting the ILL spectrometer would propagate to the reactor flux predictions. In principle, biases in both the overall normalization and the energy reconstruction (or the associated errors) might be possible, thus giving rise to the reactor \neb anomaly and the energy structure around 5 MeV, respectively. As pointed out in \cite{Dwyer} following a summation approach, the presence of a bump between 5-7 MeV in both the calculated electron and antineutrino spectra might be an indication of an artifact in the original ILL measurements rather than an effect of the conversion method.

Beyond the ILL data, the summation-derived \neb spectra from \Ureight ~(used as reference in Daya Bay, RENO and first publications by Double Chooz) might be considered a candidate to explain the energy distortion, given the large associated errors. Because of the different experimental setups, RENO reports that about 12\% of the fissions are due to \Ureight, while Daya Baya quotes only 7.6\%. As the energy structure in RENO is about 50\% larger than in Daya Bay, this might indicate that the \Ureight ~fissions are contributing to it. This isotope is indeed responsible for about 20\% of the \neb flux between 4 and 6 MeV. In \cite{Hayes15}, it has been reported that two different databases predict, within the summation scheme, bumps in the region of interest. However, the amplitude is not large enough to cover the structure observed in reactor experiments. Furthermore, the analysis in \cite{Hayes15} has not considered the work in \cite{Haag}, where the \Ureight ~aggregate spectrum has been measured. On the contrary, Double Chooz has used  \cite{Haag} to predict the \neb flux and has found that the energy bump remains, with roughly the same amplitude.  

\subsection{The conversion procedure}

The conversion technique has been reviewed extensively in the literature since the first antineutrino predictions based on the ILL data \cite{Conv}. As discussed previously, the method was recently improved in \cite{Mueller,Huber}. While the different approximations have been performed to fit the data to a number of virtual $\beta$ branches, the results have been consistent as far as energy shape and error budget are concerned (this is not the case of the overall normalization). This leaves small room for a possible issue in the technique itself. However, the ILL reactor is different to that ones currently used in the reactor experiments. The neutron flux spectra at typical pressurized water reactors (like the ones in Daya bay, Double Chooz and RENO) are harder in energy than the thermal spectrum of the ILL reactor. As highlighted in \cite{Hayes15}, this opens the possibility of epithermal neutron contributions to the \Urfive, \Punine, \Puone ~and \Ureight ~fissions, resulting in a shoulder at 5 MeV in the \neb spectrum. However, since there are not fission yield measurements for the nuclei that dominate that energy region, this hypothesis is hard to demonstrate or refute.

\subsection{Nuclear corrections}

The uncertainties quoted in \cite{Mueller} have been revisited in \cite{Hayes14}, since they lead to a significance of the reactor neutrino anomaly of about 3$\sigma$. As described in Sec.~\ref{sec:flux}, an antineutrino spectrum can be estimated from a beta spectrum if the  linear combination of operators involved in the decay, the endpoint energy and the nuclear charge are known. However, the fission $\beta$ spectra involve about 6000 decays, being forbidden about 30\% of them. This implies that some assumptions are needed when deriving the reactor \neb flux, given the limited knowledge on the structure of the forbidden transitions. Such assumptions affect eventually the error budget of the predictions in both the conversion and summation methods. In \cite{Hayes14}, it has been noticed that different treatments of the forbidden transitions (and the associated $\delta$ corrections) can lead to antineutrino spectra that differ both in shape and magnitude at about 4\%. In particular, if all forbidden decays are treated as allowed transitions, the antineutrino spectra are increased, thus leading to the reactor neutrino anomaly reported in \cite{anom}. However, this is not always the case if different approaches for the forbidden transitions are adopted. It is concluded that uncertainties in the \neb predictions are about 4\%, implying an increase of roughly a factor 2 with respect to the estimations in \cite{Mueller,Huber}. Along the same lines, the effect of first forbidden transitions on the $\beta$-decay neutrino spectra is analyzed in \cite{fang} by performing microscopic nuclear structure calculations. The authors conclude that these decays may be responsible for a fraction of the deficit of neutrinos observed in the reactor experiments. Although the works in \cite{Hayes14,fang} are addressing the issue of the reactor neutrino anomaly, the conclusions apply also to the energy structure observed around 5 MeV: the disagreement between the data and the flux prediction might be due to the forbidden transitions contributing to that energy region. 

Concerning the nuclear corrections to be applied to the forbidden transitions, three relevant points have been highlighted in \cite{Hayes15}. First, it has been noticed that several of the $\beta$ decays contributing to the bump region have a total angular momentum and parity which involve no WM correction. This increases the flux predictions with respect to \cite{Mueller,Huber,Fallot}, where this fact has not been taken into account. Second, the shape factor C(E$_e$) is not the same for all forbidden transitions: in \cite{Mueller}, the C(E$_e$) corresponding to a unique forbidden transition has been assumed for all the cases. Finally, the FS corrections for these transitions applied in the literature are always approximated. Despite these considerations, a more accurate treatment of the forbidden decays performed in \cite{Hayes15} cannot account for a significant fraction of the energy structure.

\subsection{The limitations of the summation method}

In order to get insights on the above topics, the summation method is a powerful handle as it provides an ILL-independent set of reference spectra and a tool to estimate the effect of the different nuclear corrections and assumptions. However, it is worth remarking that the errors associated to this technique are too large to establish any conclusion. Furthermore, the use of different databases can lead to somehow different conclusions. As an example, the ENDF/B.VII.1 compiled nuclear data \cite{endf} has been used in \cite{Dwyer} to derive the \neb spectra, yielding an energy bump in the antineutrino energy in the 5-7 MeV region (E$_{e}$=4-6 MeV). However, the ENDF/B.VII.1 data used in the analysis is not taking into account the new TAS measurements described in \cite{Fallot}. By using the reference spectra from \cite{Fallot} or the ENDF/B.VII.1 database combined with the new TAS measurements (hereafter updated ENDF/B.VII.1) \cite{Sonzogni}, the bump is significantly reduced. A detailed comparison of the updated ENDF/B.VII.1 and JEFF-3.11 \cite{jeff} decay libraries has been presented in \cite{Hayes15}: in the case of the JEFF-3.11-based results, the bump is totally removed. Finally, it must be noticed that the same kind of limitations of the available nuclear data arise when considering the reactor neutrino anomaly (i.e, the flux normalization). As an example, Daya Bay has measured an absolute \neb rate in good agreement with the current world average and with the ENDF/B-VII.1 prediction (i.e., no indication of anomaly). However, comparing the Daya Bay data with the JEFF-3.1.1 estimations yields a deficit in the \neb rate, thus suggesting the anomaly.

\section{Summary and discussion}
\label{sec:summary}

The three current antineutrino disappearance reactor experiments (Daya Bay, Double Chooz and RENO) have observed an energy distortion around 5 MeV in the prompt energy spectrum ($\sim$6 MeV in \neb spectrum), deviating from the predictions at a $\sim$4$\sigma$ level as an excess in the number of \neb. The structure is observed in the near detectors of Daya Bay and RENO (with baselines around 300 m) and in the far detectors of the three experiments (with baselines around 1-2 km), so it cannot be explained in terms of any standard neutrino oscillations. Given the correlation of the excess with the reactor power, the three collaborations conclude that the most likely explanation is an incompleteness or bias in the reactor flux models. Although the origin of the energy structure might not be related to the reactor antineutrino anomaly (that might be described in terms of sterile neutrino oscillations), both features reinforce the case for a revision of the current reactor flux predictions. A possible underestimation of the error budget in the reactor models, due to unknown or not well described contributions, has to be considered.    


There are two general methods to estimate the reactor fluxes as a composition of the \neb spectra from the main fissile isotopes. The most precise one, and the state-of-the-art reference for reactor experiments, is the conversion method: it relies on the $\beta$ aggregate spectra measured in the ILL, fitting the data to a set of virtual branches and converting the results into the corresponding \neb spectra. Currently, this method quotes an uncertainty of 2-3\%. The second approach is the so-called summation method: it builds the \neb spectra as the sum of each nuclide's individual $\beta$ spectrum, according to the available information in the nuclear databases. This technique typically yields an error envelope of 10-20\%. While uncertainty associated to the summation-based spectra covers the $\sim$10\% deviation observed in the experimental data around 5 MeV of the prompt energy spectrum, the error of the conversion-based spectra does not. The later method might be affected by: 1) an issue in the ILL data, 2) intrinsic limitations of the technique, and 3) nuclear effects or uncertainties not accounted for. The forbidden transitions might play a major role in the later aspect.

 
The available reactor data can be used to shed some light on the puzzle of the flux predictions. In particular, one can perform fits to the observed \neb spectrum for different sets of parameters and assumptions used in the predictions (like the number of $\beta$ branches or the treatment of forbidden transitions). The fit results can help to identify or rule out possible contributions to the prompt energy shoulder at $\sim$5 MeV. It is also possible to take the \neb spectra from Daya Bay, Double Chooz and RENO, and deconvolute them back to the corresponding $\beta$ spectra, which in turn can be confronted to the ILL data. The observation of the same structure in the $\beta$ spectra would be an indication of a bias in the ILL data. Finally, the summation method can be used to review the error on the predicted spectra by analyzing the effect of the different approximations concerning the nuclear corrections and the forbidden transitions. If it is concluded from this kind of studies than the error in the \neb spectra is larger than currently assumed (as suggested by some authors), the discrepancy between the observed data and the models would need to be reevaluated.



Despite the above considerations, the available reactor data is not enough to find out the actual origin of the energy distortion. The same applies to the current nuclear data concerning $\beta$ decays. However, the situation might improve once the near-future campaign of very short-baseline reactor experiments (meant to explore the possibility of sterile neutrino oscillations) starts delivering data. In order to rule out the possibility of an issue in the original ILL measurements, a new aggregate $\beta$ spectrum is the most direct approach. While this new measurement would be valuable, it might not be the ultimate solution to the origin of the discrepancy between the data and the models. Beyond cross-checking the ILL measurement, it is also worth exploring the limitations of the related conversion procedure, specially regarding the neutron spectra in different types of reactors. The comparison of aggregate $\beta$ spectra measured in a very thermal reactor and in a reactor with a harder neutron spectrum, as proposed in \cite{Hayes15}, would suffice to quantify the impact on the predicted \neb flux.  
 

The available nuclear data is neither capable of identifying the origin of the  energy structure by means of the summation method. In order to improve the precision and accuracy of the summation-derived predictions, new $\beta$ decay measurements are needed. In particular, improving the knowledge on the forbidden $\beta$ transitions is crucial. The summation-method itself can be used to define the list of most relevant transitions to be measured, by tagging the nuclei that contribute the most to the energy region of interest. As a matter of fact, the  main contributors have been identified in works like \cite{Dwyer,Sonzogni}. As most of the relevant transitions are first forbidden, the $\beta$ spectrum needs to be measured with high precision so the shape correction factor can be explored. As already demonstrated in the literature, the TAS technique has arose as the best option for several transitions. A new campaign of measurements will boost the capabilities of the summation method, thus becoming a major tool to resolve the nature of the reactor \neb energy structure. 

\section*{Note added in proof}

After the submission of this manuscript, the RENO collaboration has released a paper \cite{RENORS} where the observation of the energy structure is described. Although with 500 live days of data instead of 800, the paper accounts for the results presented in the Neutrino 2014 and NuTel 2015 conferences (cited in this review), including the figures of the prompt energy spectrum at both far and near detectors.

\end{document}